\documentclass[aps,prb,twocolumn,groupedaddress]{revtex4-1}
\usepackage{amsmath}
\usepackage{amssymb}
\usepackage{amsbsy}
\usepackage{amsfonts}
\usepackage{amsopn}
\usepackage{amstext}
\usepackage{graphicx}
\usepackage{amssymb}
\usepackage{amsfonts}
\usepackage{amsmath}
\usepackage{graphicx}
\usepackage[english]{babel}
\usepackage{color}
\usepackage{slashed}
\usepackage{esint}
\usepackage[dvips]{epsfig}
\usepackage[dvips]{graphicx}
\usepackage{float}
\usepackage{units}
\usepackage{textcomp}

\usepackage{hyperref}
\usepackage{slashed}

\begin{document}


\title{Electronic properties of a nanotorus}



\author{J. E. G. Silva}
\affiliation{Centro de Ci\^{e}ncias e Tecnologia, Universidade Federal do Cariri, 63048-080, Juazeiro do Norte, Cear\'{a}, Brazil}
\email{euclides.silva@ufca.edu.br}

\author{J. Furtado}
\affiliation{Centro de Ci\^{e}ncias e Tecnologia, Universidade Federal do Cariri, 63048-080, Juazeiro do Norte, Cear\'{a}, Brazil}
\email{job.furtado@ufca.edu.br}

\author{A. C. A. Ramos}
\affiliation{Centro de Ci\^{e}ncias e Tecnologia, Universidade Federal do Cariri, 63048-080, Juazeiro do Norte, Cear\'{a}, Brazil}
\email{antonio.ramos@ufca.edu.br}

\date{\today}

\begin{abstract}
In this paper we study the properties of an electron trapped on a torus surface. We consider the influence of surface curvature on the spectrum and the behaviour of the wave function. In addition, the effects of external electric and magnetic fields were explored. Two toroid configurations were considered, namely, a torus-like configuration and a ring-like configuration both in agreement with experimental data. We also obtain the bound states and discuss the role of geometry on the Landau levels.
\end{abstract}

\pacs{}

\maketitle

\section{Introduction}

The striking properties of two dimensional nanostructures, such as nanotubes \cite{berber}, graphene \cite{geim1, geim2, katsnelson} and phosphorene \cite{carvalho} have sparked many theoretical and experimental research. A single layer graphene exhibits no gap in the conductance band and an electron on its surface behaves as a massless particle \cite{katsnelson}. On the other hande, a bilayer graphene possesses a quadratic dispersion relation leading to a gap in the conductance band, and interesting applications to electronics \cite{katsnelson}.

The influence of the surface geometry on the electrical properties is a topic analysed since the early days of the quantum mechanics \cite{dirac,dewitt,chang}. By defining tangent and normal coordinates and taking the limit where the surface width vanishes, a curvature-dependent potential known as da Costa potential emerges. This squeezing method also enables the inclusion of external fields \cite{luiz}, spin in a non-relativistic equation \cite{ferrari} and in the Dirac equation \cite{wang}.

New electronic devices can be produced based on curved graphene structures. Graphene strips in a helical exhibit chiral properties \cite{dandoloff,atanasov1,atanasov2} known as chiraltronics, whereas M\"{o}bius-strip graphene behaves as topological insulator material \cite{guo}. The effects of ripples \cite{juan} and corrugated \cite{atanasov3} surfaces upon electrons can also be described by geometric interactions. Curved graphene sheet also gives rise to effective interactions such as pseudomagnetic fields \cite{guinea}. A bridge connecting two parallel layers of graphene was proposed using a nanotube \cite{gonzalez, pincak} and a catenoid surface \cite{dandoloff2, dandoloff3, euclides}.  

Another interesting geometry studied in the last years is the torus surface. In the theoretical framework, the curvature-induced bound-state eigenvalues and eigenfunctions were calculated for a particle constrained to move on a torus surface considering that such particle is governed by the Schr\"{o}dinger equation \cite{mario}. Analytical solution for the Pauli equation for a charged spin $1/2$ particle moving along a toroidal surface was found in \cite{alexandre}. Magnetic properties of toroidal nanomagnets were studied in \cite{vojkovic}. Bending carbon nanotubes to produce nanotorus are considered theoretically and experimentally in \cite{liu, sano}. Besides, the first experimental observation of carbon toroids was reported in \cite{liu2} and since then several different experimental techniques were used to produce such nanostructure \cite{Ahlskog, wang, martel}. In this paper we explore the effects of the curvature-dependent potential and the external fields upon an electron confined on a torus surface. In absence of the external fields, the curvature and the azimuthal quantum number control the regions and the profile of the bound states. We considered both a large torus radius and a ring-like configurations, relating to experimental data. The energy spectrum and its behaviour with the geometry and the electric and magnetic fields were also investigated.

This paper is organized as follows: in section II we present the system itself and we discuss in details the geometric potential induced by the torus surface as well as the influence of external electric and magnetic fields. In section III we obtain numerically the bound states and the probability densities for two torus configurations, toroid and ring-type toroid, using the effective mass method. In section IV we highlight our conclusions and future perspectives.

\section{Electron on a torus surface}
\label{section1}

\begin{figure}[h!]
\begin{center}
\includegraphics[scale=0.25]{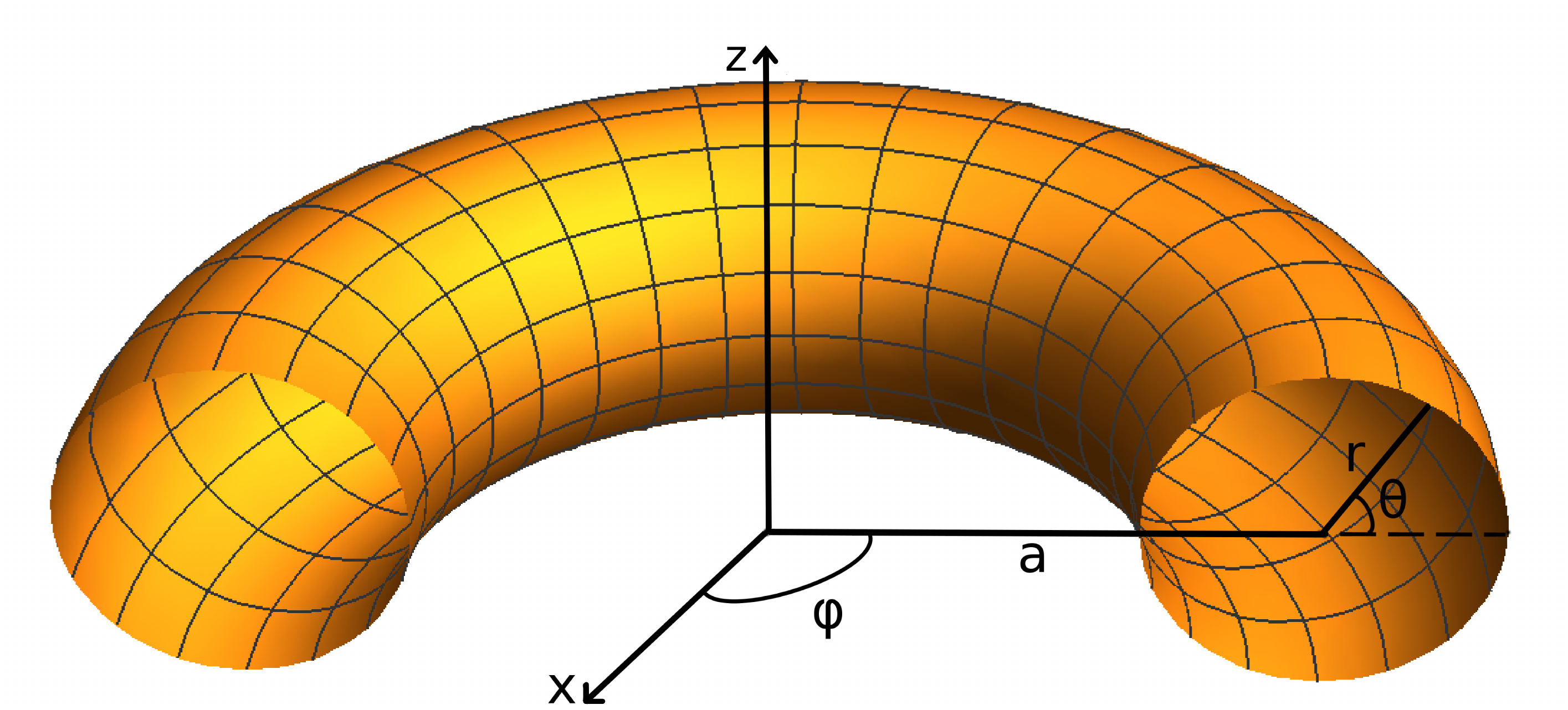}
\caption{Torus section.}
\label{throat}
\end{center}
\end{figure}

In this section we introduce the geometry and the dynamics of the electron on a torus surface, which can be realized by connecting the two ends of a carbon nanotube. The symmetry about the $z$ axis and the geometric parameters $a$ and $r$ are shown in the Fig.(\ref{throat}).

Consider a Hamiltonian quadratic in the momentum of the form \cite{ferrari,wang}
\begin{equation}
\label{totalhamiltonian}
\hat{H}=\frac{1}{2m^{*}}g^{ij}\hat{P}_i \hat{P}_j + V_{e} + V_g,
\end{equation}
where $\hat{P}_i :=-i \hbar \nabla_i - e A_i$ is the momentum operator of the electron minimally coupled to the magnetic field, $V_e$ is the electrostatic potential, $V_g$ is a confining potential which constrains the electron on the surface, and $m^{*}$ is the effective mass of the electron on a toroid made of graphene.

The Hamiltonian \eqref{totalhamiltonian} is invariant upon change of coordinates of the surface and the metric $ds^2=g_{ij}(x)dx^{i}dx^{j}$ is the induced metric on the torus. Throughout this work the index $i,j=\{1,2\}$ and stand label the torus coordinates.

We adopt the torus coordinates as
\begin{equation}
\label{cilindricalcoordinates}
\vec{r}(\theta,\phi)=(a+r\cos\theta)\cos\phi\hat{i} +(a+r\cos\theta)\sin\phi\hat{j} + r\sin\theta\hat{k},
\end{equation}
where $a$ is the radius of the torus itself, from the origin of the coordinate system to the center of the nanotube, while $r$ is the nanotube radius. The coordinates $\theta\in[-\pi,\pi)$ and $\phi\in[0,2\pi)$ cover the whole torus.

\begin{figure}[h]
\begin{center}
\includegraphics[scale=0.5]{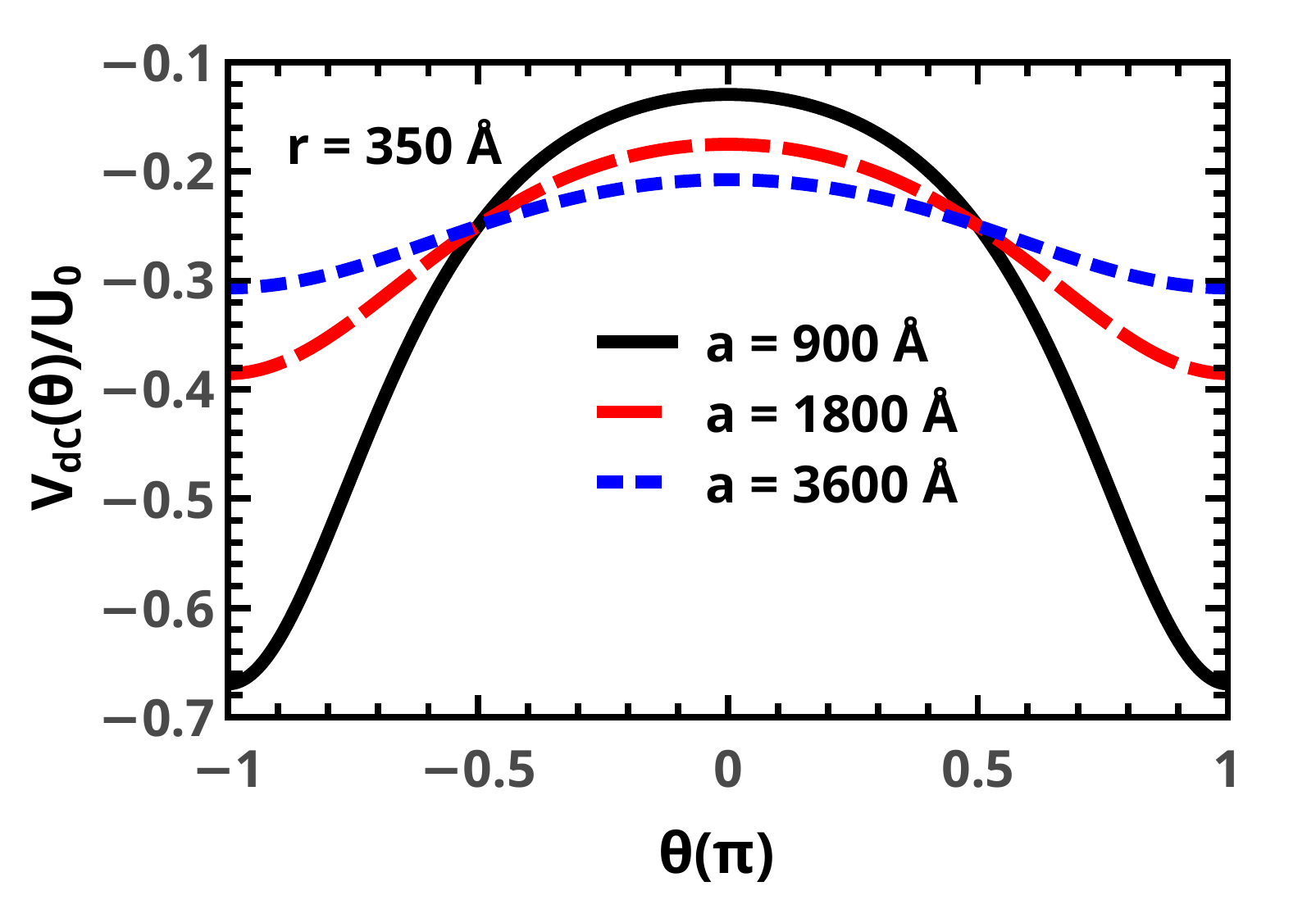}
\caption{Geometric da Costa potential, $V_{dC}$, on torus for $r=350$\AA, for some values of $a$.}
\label{dacostapotential}
\end{center}
\end{figure}

In this coordinate system, the interval reads
\begin{equation}
ds^{2}=r^2d\theta^2 + (a+r\cos\theta)^2d\phi^2,
\end{equation}
and then, the diagonal components of the induced metric tensor on the torus are $g_{\theta\theta}=r^2$ and $g_{\phi\phi}=(a+r\cos\theta)^2$. The nonvanishing components of the Christoffel connection are
\begin{eqnarray}
\Gamma_{\phi\phi}^{\theta}&=&\frac{(a+r\cos\theta)\sin\theta}{r}\\
\Gamma_{\theta\phi}^{\phi}&=&-\frac{r\sin\theta}{a+r\cos\theta}.
\end{eqnarray}
Thus, the spinless stationary Schr\"{o}dinger equation is written as
\begin{eqnarray}
\nonumber &&\frac{1}{2m^{*}}g^{ij}\left\{-\hbar^2\left[\partial_i\partial_j\psi-\Gamma_{ij}^k\partial_k\psi\right]\right.+ie\hbar(\nabla_iA_j)\psi\\
&+&2ie\hbar A_i\partial_j\psi+\left.e^2A_{i}A_{j}\psi+V_{dC}\psi\right\} = \epsilon\psi.
\end{eqnarray}
We consider the geometrical da Costa potential $V_{dC} = - \frac{\hbar^2}{2m^{*}}(H^2-K)$, where $H$ is the mean curvature and $K$ is the gaussian curvature (\cite{costa}). In the torus, the da Costa potential reads
\begin{equation}\label{e1}
V_{dC}(\theta) =-\frac{\hbar^2}{8m^{*}}\frac{a^2}{r^2(a + r\cos\theta)^2}.
\end{equation}
 where $m^{*}=0.3m_{0}$, and $m_{0}$ is the electron rest mass, this value is used throughout work \cite{Haddon}. The Fig.\eqref{dacostapotential} shows the potential, given by the Eq.(\ref{e1}), for $r=350$\AA. In the plot, we also consider $U_{0}=\hbar^{2}/2m^{*}r^{2}=0.10496$ meV. The solid black, dashed red and dotted blue lines correspond to $a=900,1800$ and $3600$\AA, respectively \cite{Ahlskog}. Here, $\theta=-\pi$ and $\theta=\pi$ are equal due to periodicity of the system. The da Costa potential, for the geometry of the toroid, exhibits a potential well with its minimum at $\theta=\pi$. Moreover, the well decreases its depth as the radius of the toroid, $a$, increases. Therefore, as we can see in Fig.\eqref{dacostapotential}, when the ratio $a/r$ becomes larger (ring limit) the net effect of da Costa's potential becomes negligible. The chosen torus dimensions are in agreement with the experimental values of real nanotorus produced \cite{Ahlskog}. 

In the next subsection we will focus our attention in the behaviour of an electron in absence of external electromagnetic fields moving on the torus surface.

\subsection{Curvature effects}
\label{freelectron}

Let us study the properties of the wave function in absence of the electric and magnetic fields.
The Schr\"{o}dinger equation becomes
\begin{eqnarray}
-\frac{\hbar^2}{2m^{*}}\left\{g^{ij}\left[\partial_i\partial_j\psi-\Gamma_{ij}^k\partial_k\psi\right]+V_{dC}\psi\right\} = \epsilon\psi.
\end{eqnarray}
which can also be written explicitly in terms of the torus parameters as
\begin{eqnarray}
\nonumber-\frac{\hbar^2}{2m^{*}}\Bigg[\frac{1}{r^2}\psi_{\theta\theta}&+&\frac{1}{(a+r\cos\theta)^2}\psi_{\phi\phi}-\frac{\sin\theta}{r(a+r\cos\theta)}\psi_{\theta}\nonumber\\
&-&\frac{a^2}{4r^2(a + r\cos\theta)^2}\psi\Bigg] = \epsilon\psi.
\end{eqnarray}
In the previous equation $\psi_j=\partial\psi/\partial x^j$. Considering $V_{dc}=0$ and the axial symmetry ($\phi\rightarrow-\phi$), so that, $\psi(\theta,\phi)=\Phi(\theta)e^{im\phi}$, we can rewrite the Schr\"{o}dinger equation as
\begin{eqnarray}\label{s1}
-\frac{\hbar^2}{2m^{*}}\Bigg[\frac{1}{r^2}\frac{\partial^2\Phi}{\partial\theta^2}&-&\frac{\sin\theta}{r(a+r\cos\theta)}\frac{\partial\Phi}{\partial\theta}\nonumber\\
&-&\frac{m^2}{(a+r\cos\theta)^2}\Phi\Bigg]= \epsilon\Phi.
\end{eqnarray}

Note that the Eq.(\ref{s1}) exhibits parity and time-reversal symmetry invariance, as a result of the torus geometric symmetries that we can see from the acting of parity $\mathcal{P}$ and time reversal $\mathcal{T}$ operators. Besides, the first order derivative term leads to a hamiltonian non-hermitean, since
\begin{eqnarray}
\Bigg[\frac{i\hbar}{2m^{*}}\frac{\sin\theta}{(a+r\cos\theta)}\hat{P}_{\theta}\Bigg]^{\dagger}&=&-\frac{\hbar^2}{2m^{*} r}\Bigg[\frac{\cos\theta}{(a+r\cos\theta)}\nonumber\\
    &+&\frac{r\sin^2\theta}{(a+r\cos\theta)^2}\Bigg]\nonumber\\
    &-&\frac{i\hbar}{2m^{*}}\frac{\sin\theta}{(a+r\cos\theta)}\hat{P}_{\theta},
\end{eqnarray}
where $\hat{P}_{\theta}=-i\hbar r^{-1}\partial_{\theta}$ and $\hat{\theta}$ are indeed Dirac hermitean. The non-Hermiticity of the free electron Hamiltonian is not a serious issue, since space-time reflection symmetry is preserved, the spectrum of the eigenvalues of the Hamiltonian is completely real \cite{Bender, Bender2}. Besides, there is an Hermitean equivalent Hamiltonian \cite{Bender, Bender2} that can be found by a simple changing of variables. Considering now that $\Phi(\theta)=y(\theta)\chi(\theta)$, where $\chi(\theta)$ obeys the Schr\"{o}dinger equation. Applying this transformation into (\ref{s1}) and choosing $y(\theta)$ in order to eliminate $\chi'(\theta)$ we reach at
\begin{eqnarray}\label{e12}
    -\frac{\hbar^2}{2m^{*}r^2}\chi''+V_{ind}(\theta)\chi=\epsilon\chi,
\end{eqnarray}
with the inducted potential $V_{ind}(\theta)$ being 
\begin{eqnarray}
V_{ind}(\theta)=\frac{\hbar^2}{2m^{*}r^2(a+r\cos\theta)^2}\times \nonumber\\ \left[r^2m^2+\frac{r^2\sin^2\theta}{4}-\frac{r(a\cos\theta+r)}{2}\right].
\end{eqnarray}
The inclusion of the da Costa potential yields the effective potential given by
\begin{eqnarray}\label{poteg}
    V_{eff}(\theta)=\frac{\hbar^2}{2m^{*}r^{2}(a + r\cos\theta)^2}\times \nonumber \\ \left[-\frac{a^{2}}{4}+r^2m^2+\frac{r^2\sin^2\theta}{4}-\frac{r(a\cos\theta+r)}{2}\right].
\end{eqnarray}
Therefore, the behaviour of an electron confined on a torus surface can be described by an Hermitean Hamiltonian with an effective one-dimensional potential written as (\ref{poteg}). In addition, $|\chi|^2$ represents a probability density, for Eq. (\ref{e12}) satisfies the continuity equation. On the other hand, the first-order derivative in Eq. (\ref{s1}) avoids the density $|\Phi|^2$ to satisfy the continuity equation.

\begin{figure*}[h]
\begin{center}
\includegraphics[scale=0.5]{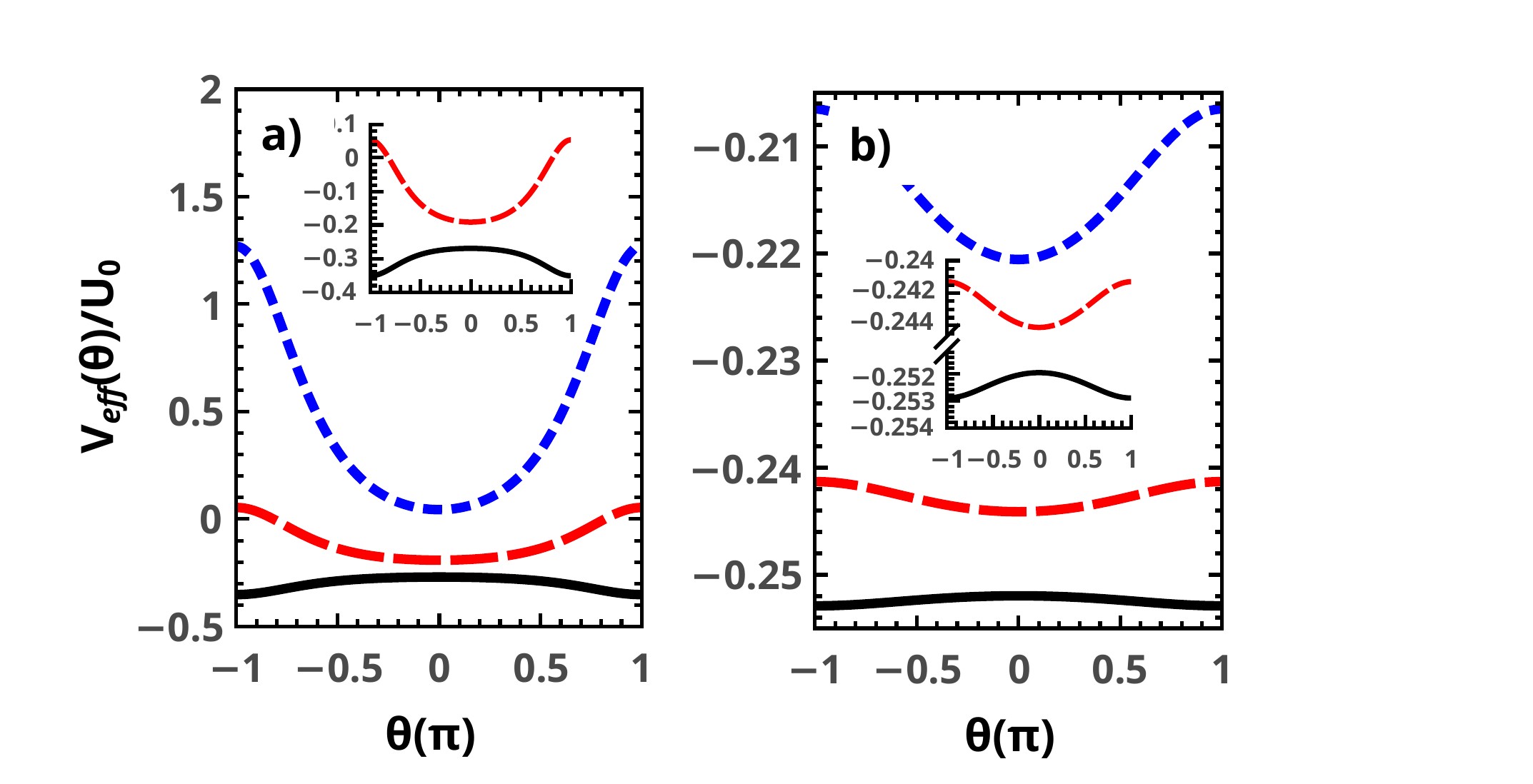}
\caption{The effective potential, given by the Eq.(\ref{poteg}), for $r=350$\AA. The figures a) for $a=900$\AA, and b) for $a=3600$\AA. The solid black line represents the potential for $m=0$, the other lines represent the potential for $m\neq 0$. The inserts in the figures highlight the effective potentials for $m=0$ and $m=1$.}
\label{geometricpotential1}
\end{center}
\end{figure*}

The Figs.(\ref{geometricpotential1}) a) and b) show the effective geometric potential, given by the Eq.(\ref{poteg}), for the first three values of angular momentum, $m$.  The solid black, dashed red and dotted blue lines correspond to $m=0,1$ and $2$, respectively. In our calculations experimental data were used again to set the dimensions of the nanotorus, so that we choose $r=350$\AA, $a=900$\AA and $3600$\AA. As we can see in the Figs.(\ref{geometricpotential1}) a) and b) that the effective potential, given by the geometry of the toroid, exhibits an interesting behavior given by the orbital angular momentum $m$.

When $m\neq 0$ the minimum of the effective potential is located at $\theta=0$, and for larger $m$ the depth of the potential well is greater. On the other hand, when $m=0$ the minimum of the effective potential is located at $\theta=\pi$. This effect is highlighted by the inserts of the Figs.(\ref{geometricpotential1}) a) and b), for $a=900$ and $3600$\AA, respectively. 

It is worth noting that the toroid, with the dimensions $a=3600$ and $r=350$ \AA, when observed by a scanning electron microscope, it no longer has the characteristic of a toroid to have the characteristics of a ring, or ring-type toroid. And under these conditions, the effective geometric potential is reduced tremendously, as shown in the Fig.(\ref{geometricpotential1}) b) when compared to the Fig.(\ref{geometricpotential1}) a), which represents, in fact, a toroid. So when we refer to the effective potential of the toroid type, we are referring to the toroid with the dimensions of $a=900$ and $r = 350 $ A. And when we refer to the effective potential of the ring type, we are referring to the toroid with the dimensions $a=3600$ and $r=350$ \AA.


\subsection{External Electric Field}

In this section, we consider the effect of a constant external electric field on an electron on the torus. Considering the electric field pointing in the positive $z$ direction, i.e., $\vec{E}=E\vec{k}$, the projection of the electric field on the torus gives rise to an electric potential of the form:
\begin{eqnarray}\label{a4e}
    V_{e}(\theta)=-eEr\sin\theta.
\end{eqnarray}
Note that the electric potential breaks the parity symmetry of the system, i.e., $\mathcal{P}V_e(\theta) \mathcal{P}=-V_e(\theta)$. 
Such break of parity promotes a displacement of the electron's confining region which depends on the intensity of the electric field. Therefore, the effective potential of an electron on the surface of the toroid under the action of a constant electric field is given by
\begin{eqnarray}\label{a2e}
    V_{eff}(\theta)&=&\frac{\hbar^2}{2m^{*}r^{2}(a + r\cos\theta)^2}\Bigg[-\frac{a^{2}}{4}+r^2m^2\nonumber\\
    &+&\frac{r^2\sin^2\theta}{4}-\frac{r(a\cos\theta+r)}{2}\Bigg] -eEr\sin\theta.
\end{eqnarray}

\begin{figure*}[h!]
\begin{center}
\includegraphics[scale=0.5]{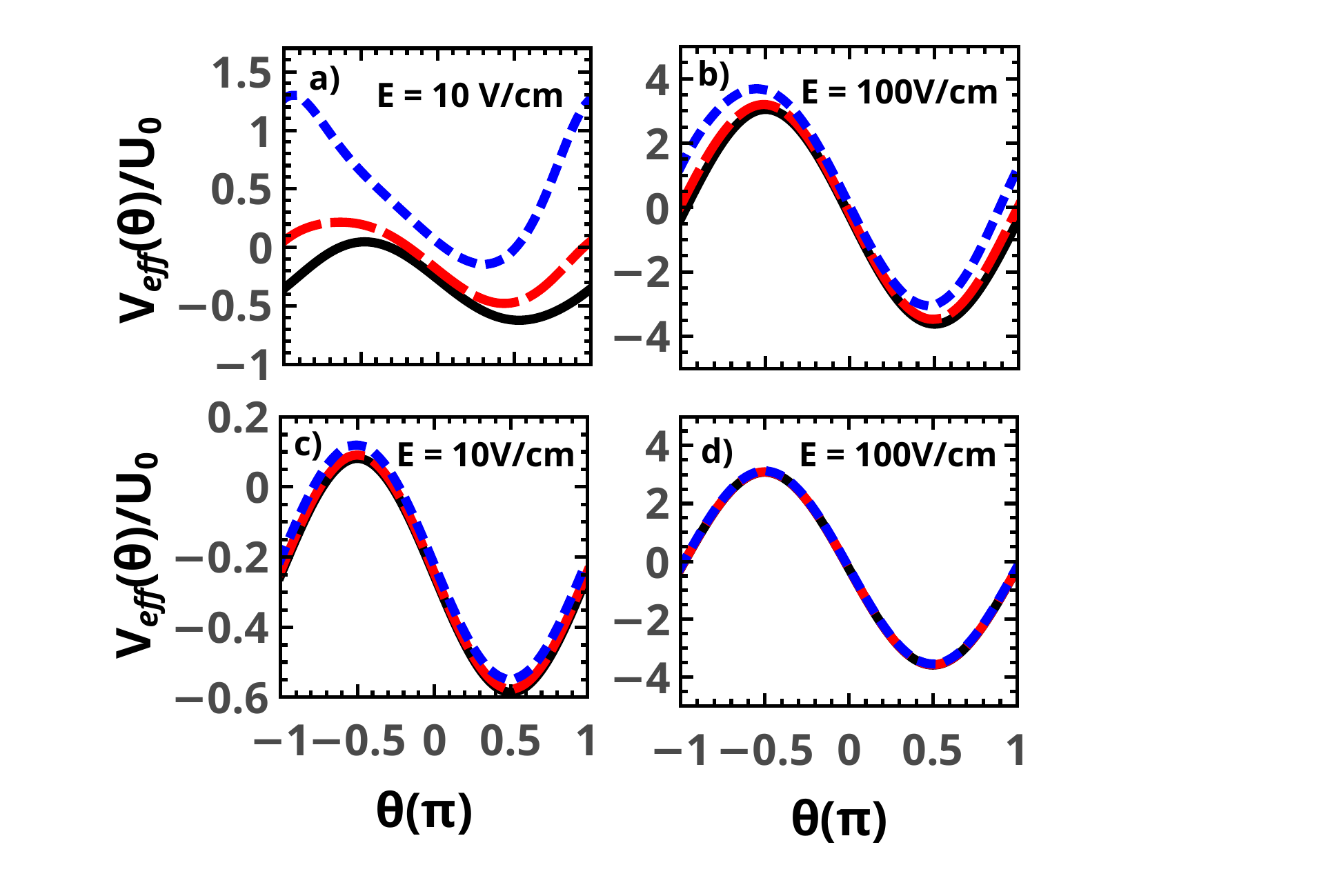}
\caption{The effective potential, given by the Eq.(\ref{a2e}), for $r=350$ \AA. In a) $E=10$ V/cm and b) $E=100$ V/cm for $a=900$\AA. In c) $E=10$ V/cm and d) $E=100$ V/cm for $a=3600$\AA. The solid black, the dashed red and the dotted blue lines correspond to $m = 0, 1$ and $2$, respectively.}
\label{electricpotential}
\end{center}
\end{figure*}
The Fig.(\ref{electricpotential}) a) shows several effective potentials for some values of angular momentum, $m$, for $r=350$\AA, $a=900$\AA\ and $E=10$ V/cm. Comparing the Fig.(\ref{geometricpotential1}) a) with the Fig.(\ref{electricpotential}) a) we note that the application of an electric field, although not very intense, changes drastically the minimum of the effective potential, mainly for the three potentials with the lowest orbital angular momentum, $m=0, 1$ and $2$. We can see that the minimum of the effective potential, for $m=0$, is shifted from $\theta=\pi$ to $\theta=\pi/2$. While the minimum of the potential for $m=1$ and $2$ changes, to the right, from $\theta=0$ to some value of $\theta$ between $0$ and $\pi/2$. For larger $m$ the effect of the electric field, $E=10$ V/cm, is less expressive. This effect described above for $E=10$ V/cm is accentuated for more intense fields, $E=100$ V/cm, as is shown in the Fig.(\ref{electricpotential}) b). 

For Figs.(\ref{electricpotential}) c) and d), we find that the effect of the electric field is more significant because the geometric effect is reduced drastically. We also observe that in addition to the electric field producing the displacement of the minimum potentials, they can also make them deeper.



\subsection{External Magnetic Field}

Considering now the action of a constant external magnetic field on the system, we have a vector potential written as $\vec{A}=\frac{1}{2}\vec{B}\times\vec{r}=\frac{1}{2}B(a+r\cos\theta)\hat{\phi}$. The effective magnetic potential becomes:
\begin{eqnarray}\label{magpot}
    V_{m}=\frac{e^2B^2}{8m^{*}}(a+r\cos\theta)^2-\frac{me\hbar B}{2m^{*}}
\end{eqnarray}
Therefore, with the magnetic field, the effective potential becomes
\begin{eqnarray}\label{a3e}
    V_{eff}(\theta)=\frac{\hbar^2}{2m^{*}r^{2}(a + r\cos\theta)^2}\times \nonumber \\ \left[-\frac{a^{2}}{4}+r^2m^2+\frac{r^2\sin^2\theta}{4}-\frac{r(a\cos\theta+r)}{2}\right] \nonumber \\-eEr\sin\theta+\frac{e^2B^2}{8m^{*}}(a+r\cos\theta)^2-\frac{me\hbar B}{2m^{*}}
\end{eqnarray}

Let us highlight here that the last term in eq.(\ref{a3e}) breaks chiral symmetry. Hence the presence of an external magnetic field will display a difference between the potentials with $m>0$ and $m<0$.

\begin{figure*}[h!]
\begin{center}
\includegraphics[scale=0.5]{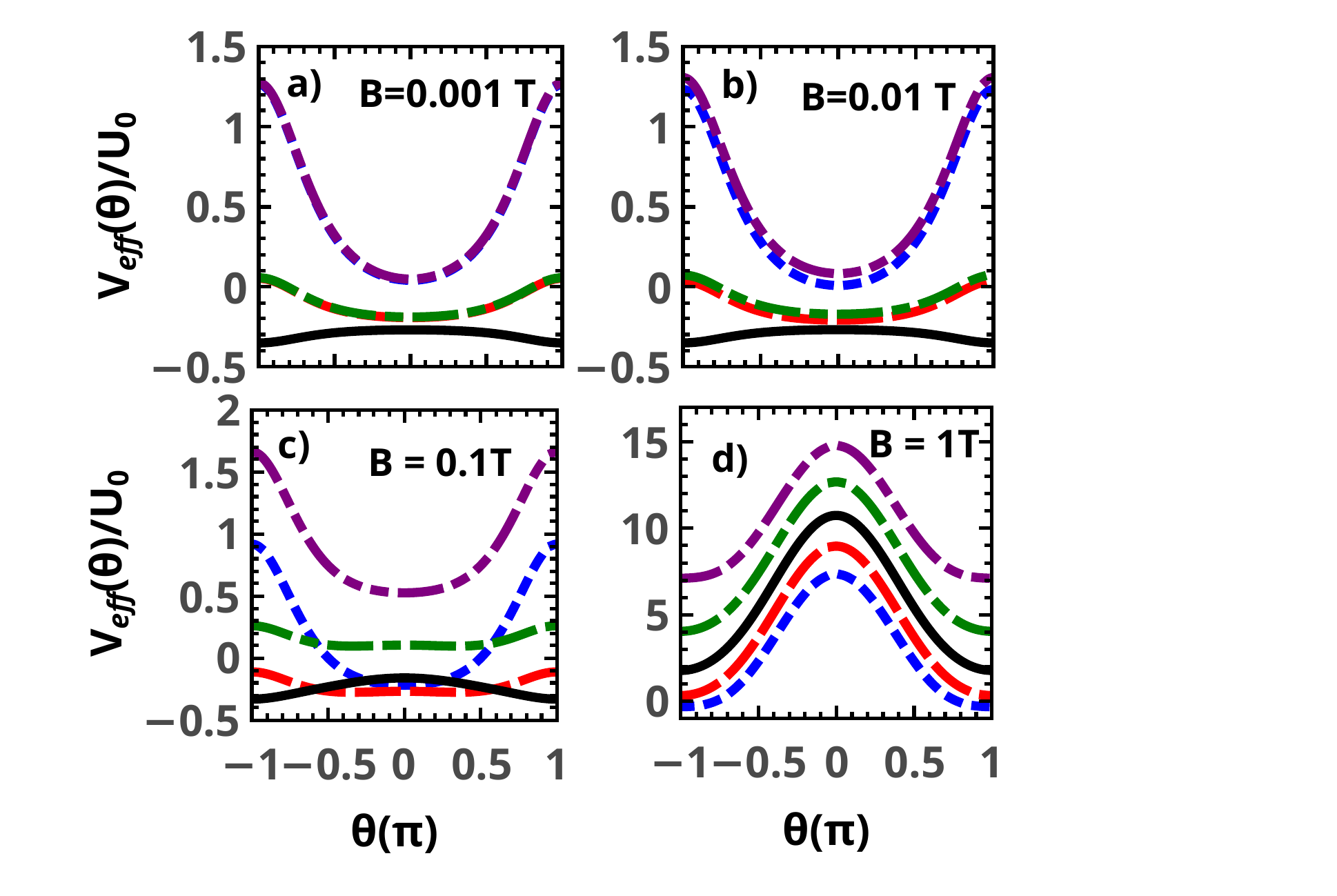}
\caption{The effective potential, given by the Eq.\ref{a3e}, for $E=0$ V/cm. The torus configuration is $r=350$\AA and $a=900$\AA. The dash-dot-dotted purple, dash-dotted green, solid black, dashed red and dotted blue lines correspond to the potentials for $m=-2,-1,0,1$ and $2$, respectively.}
\label{magneticpotential1}
\end{center}
\end{figure*}

The Figs.(\ref{magneticpotential1}) a), b), c) and d) show us the behaviour of the system with $r=350$\AA \,\, $a=900$\AA \,\, and $E=0$ V/cm, under the action of four intensities of magnetic field, namely, $B=0.001$ T, $B=0.01$ T, $B=0.1$ T and $B=1$ T, respectively. In all the plots considering magnetic field, the break of chirality is evident for $B>0.001$ T. The Fig.(\ref{magneticpotential1}) a) shows the effective potentials curves, for $B=0.001$ T,  the geometric potential is entirely dominant. We observe that the curves of the potentials for $m=-2$ and $m=2$ are overlapping. The same occurs with the potentials for $m=-1$ and $m=1$. For an intensity of $B=0.01$ T, Fig. (\ref{magneticpotential1}) b), the geometric potential is still dominant, however the overlapping of the potential curves for $m=2$ and $m=-2$ begins to be removed, the same occurs for $m=1$ and $m=-1$. This effect is what we call the chirality break. 

In Fig.(\ref{magneticpotential1}) c), for $B=0.1$ T, the effect of the magnetic field becomes slightly more relevant and the break of chirality is accentuated for $m=\pm1$ and $m=\pm2$, besides making potentials deeper.

When we compare the results obtained for the effective potentials for $B=0$ T, shown by the Fig.(\ref{geometricpotential1}) a), and $B=1$ T, shown by the Fig.(\ref{magneticpotential1}) d), we see that the minimum of the effective potentials, for $m=1$ and $m=2$, are shifted from $\theta=0$ to $\theta=\pi$, in addition we observe that these potential wells become deeper.

\begin{figure*}[h!]
\begin{center}
\includegraphics[scale=0.7]{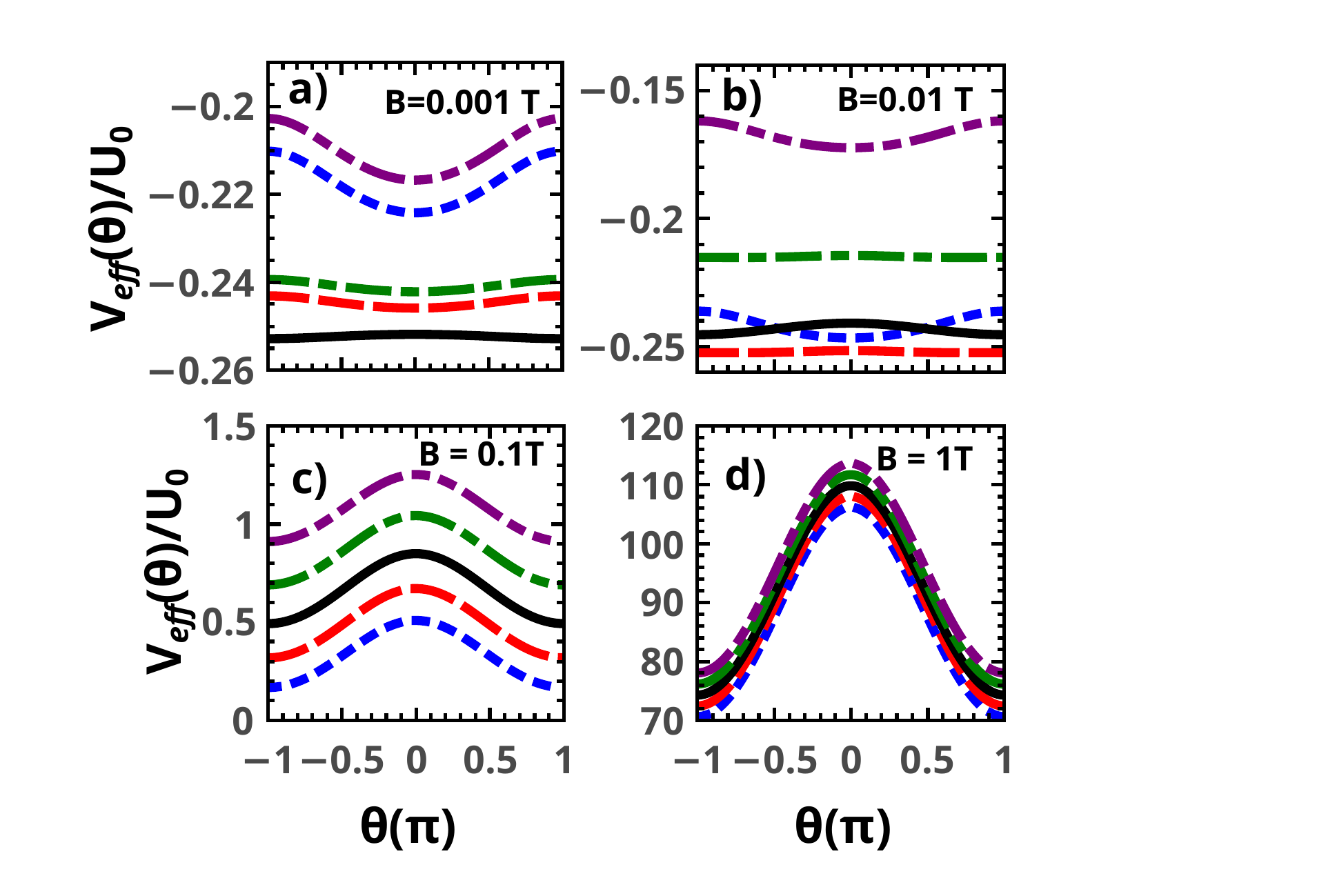}
\caption{The effective potential, given by the Eq.\ref{a3e}, for $E=0$ V/cm. The torus configuration is $r=350$\AA \,\,and $a=3600$\AA. The dash-dot-dotted purple, dash-dotted green, solid black, dashed red and dotted blue lines correspond to the potentials for $m=-2,-1,0,1$ and $2$, respectively.}
\label{magneticpotential2}
\end{center}
\end{figure*}

We also investigate the effect of the magnetic field on a ring-type toroid , with dimensions $a=3600$ \AA \,\ and  $r=350$ \AA. The results are shown in the Fig.(\ref{magneticpotential2}). We observe that the effective potentials behave qualitatively in a similar way. But, there is a quantitative change due to the decrease in the effect of the geometric confinement, already highlighted in the text. For these dimensions, of the ring-type toroid, we observe that the effect of the magnetic field increases the depth of the potential well, given by Eq.(\ref{a3e}), in the region of $\theta=\pi$, shown in the Fig.(\ref{magneticpotential2}) d), whe we compare with the case of the toroid, Fig.(\ref{magneticpotential1}) d).

\begin{figure*}[h!]
\begin{center}
\includegraphics[scale=0.7]{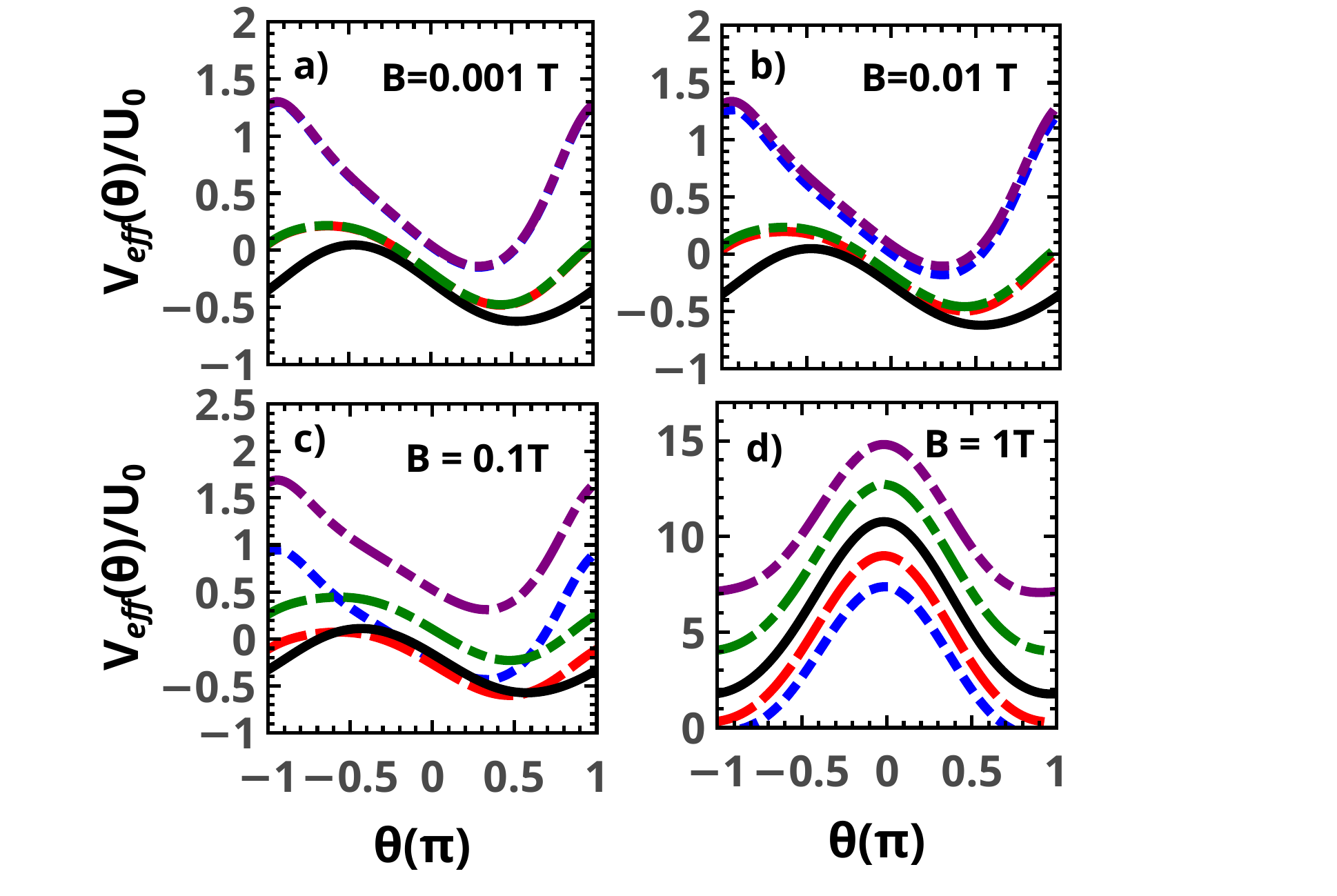}
\caption{The effective potential, given by the Eq.(\ref{a3e}), for $E=10$ V/cm. The torus configuration is $r=350$\AA \,\,and $a=900$\AA. The dash-dot-dotted purple, dash-dotted green, solid black, dashed red and dotted blue lines correspond to the potentials for $m=-2,-1,0,1$ and $2$, respectively.}
\label{electricmagnetic1}
\end{center}
\end{figure*}

In the Figs.(\ref{electricmagnetic1}) a), b), c) and d), show the effect of electric and magnetic field combined on the toroid. In these figures the intensity of the electric field is fixed, $E=10$ V/cm, and we vary from figures a) to d) the intensity of the magnetic field. We see that in the Fig.(\ref{electricmagnetic1}) a), b) and c) the effect of the electric field dominates over the geometric and the magnetic, mainly for effective potentials with $m=0$ and $\pm1$. We also note that in the Fig.(\ref{electricmagnetic1}) d), for $B=1$ T, the effect of the magnetic field dominates over other.



\begin{figure*}[h!]
\begin{center}
\includegraphics[scale=0.7]{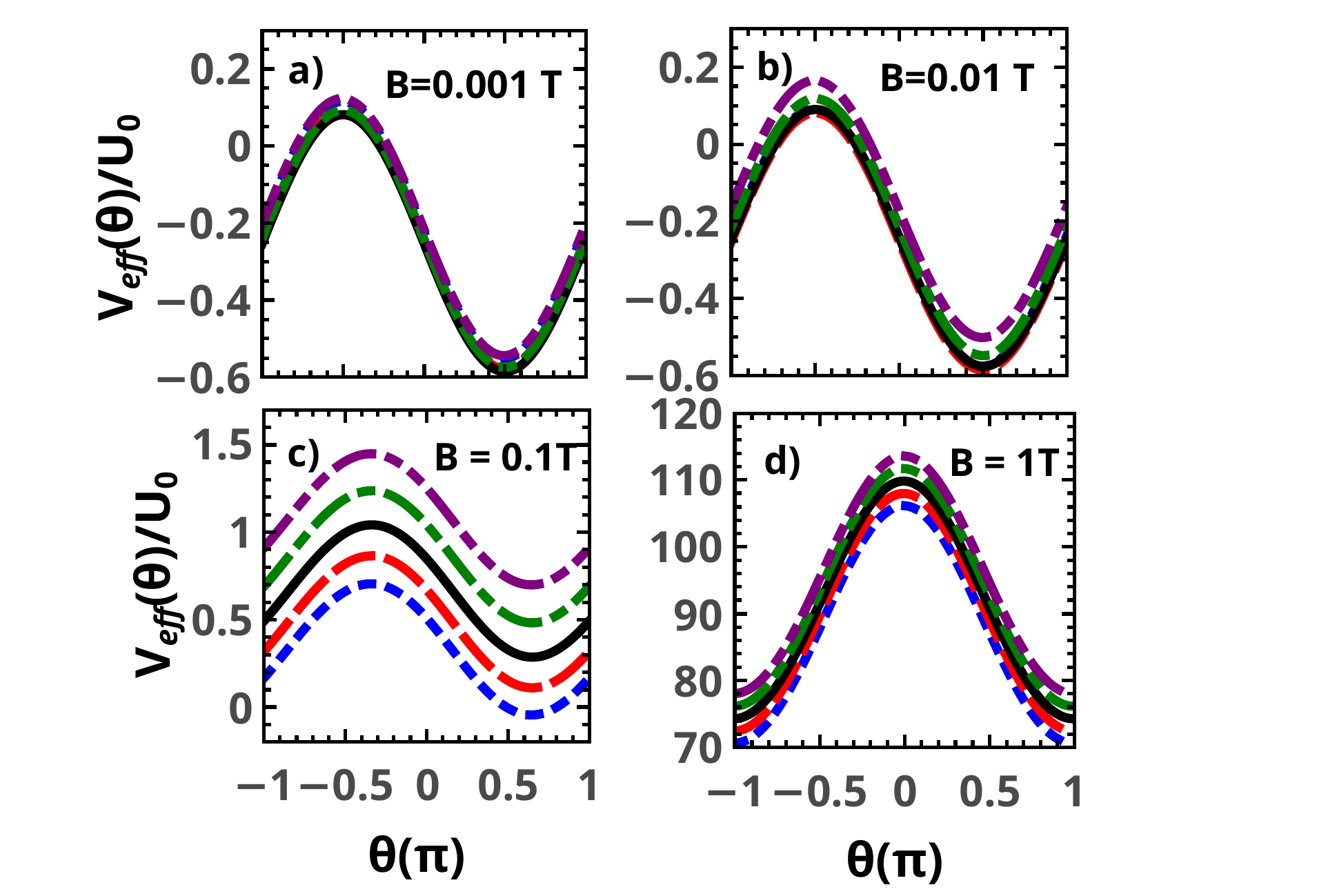}
\caption{The effective potential, given by the Eq.(\ref{a3e}), for $E=10$ V/cm. The torus configuration is $r=350$\AA \,\,and $a=3600$\AA. The dash-dot-dotted purple, dash-dotted green, solid black, dashed red and dotted blue lines correspond to the potentials for $m=-2,-1,0,1$ and $2$, respectively}
\label{electricmagnetic2}
\end{center}
\end{figure*}
In the Figs.(\ref{electricmagnetic2}), the ring-type toroid is under also the action of electric and magnetic fields. For this geometry the effects of the fields are more significant than the geometric. We can see that up to $B=0.1$ T, for $E=10$ V/cm, the electric field is dominant, for various values of $m$, and that the minimum of the effective potential is dislocate to $\theta=\pi/2$, and becomes deeper for $B=0.1$ T. The effect of the magnetic becomes dominant for $B=1$ T, and the minimum of the potential is dislocated to $\theta=\pi$.   
By comparing the effective potential of the toroid with ring-type toroid, for $E=10$ V/cm and $B=1$ T, the latter to see that the effective potential is deeper, as shown in the Figs.(\ref{electricmagnetic1}) d) and (\ref{electricmagnetic2}) d).

Through these figures we can clearly see that the competition between the electric and magnetic fields, dominates over the confinement produced by the geometry of the toroid, mainly of the ring-type toroid. This facility in manipulating the effective potential due to external effects opens up a vast technological application for this system.

\section{Bound States}

In this section we present the energy levels and the probability density of an electron on the graphene toric surface, under the action of electric and magnetic fields. 

The energy levels and their probability densities are obtained by solving the equation Schrödinger given by 
 \begin{eqnarray}\label{a4e}
    -\frac{\hbar^2}{2m^{*}r^2}\frac{d^{2}\chi(\theta)}{d\theta^{2}}+V_{eff}(\theta)\chi(\theta)=\epsilon\chi(\theta).
\end{eqnarray}

\begin{figure*}[h!]
\begin{center}
\includegraphics[scale=0.6]{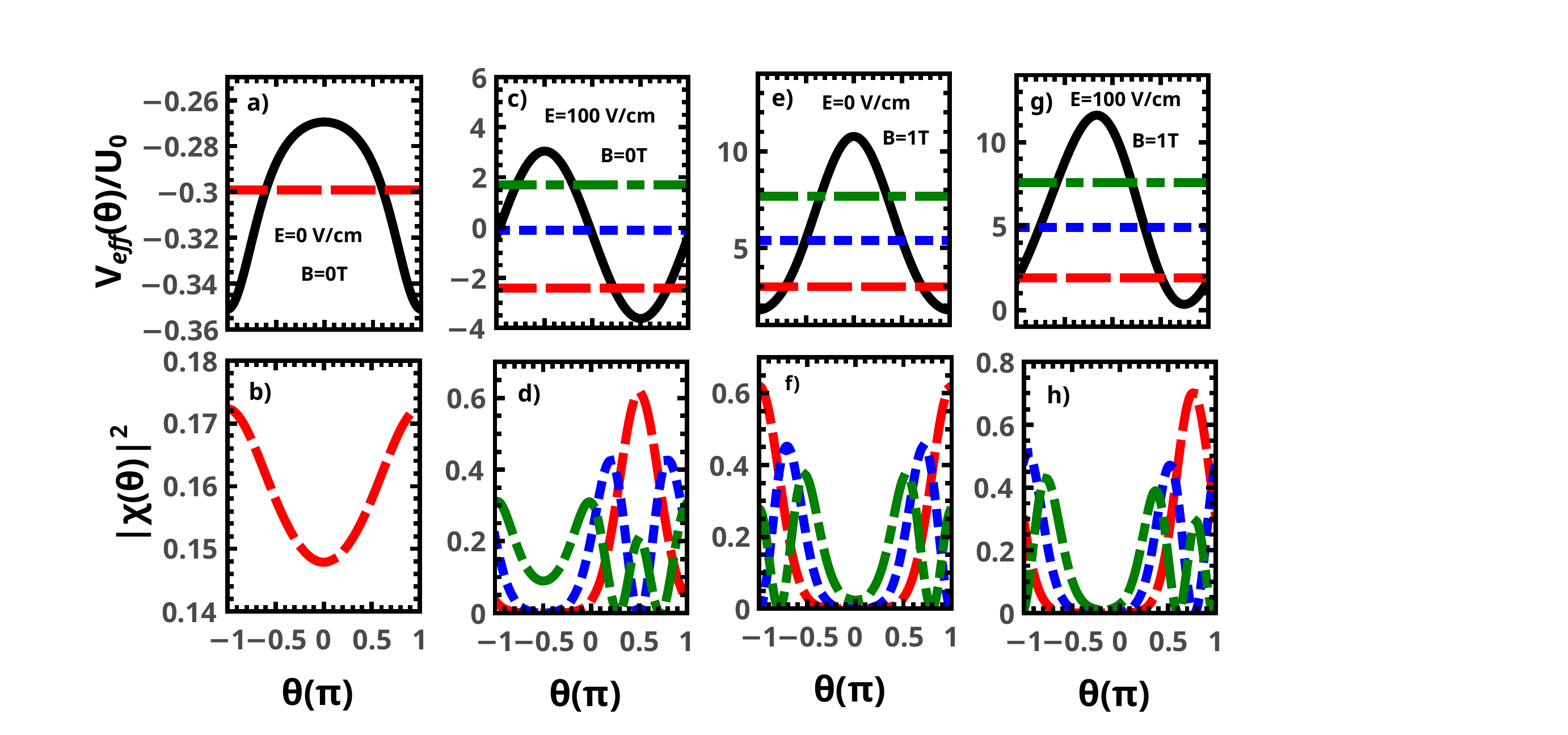}
\caption{Bounds states and the their probability densities for a torus configuration with $r=350$\AA \,\,and $a=900$\AA. The solid black line represents the effective potential, for $m=0$. The dashed red, the dotted blue and the dash-dotted green lines correspond to the first, the second and the third energy levels and their probability densities, respectively.}
\label{BS1}
\end{center}
\end{figure*}

We solve the Eq.(\ref{a4e}), using the Eq.(\ref{a3e}), for $m=0$, $a=900$ \AA \,\ and $r=350$ \AA, and we obtained the energy levels and probability densities, for some values of the electric, $E$, and magnetic, $B$, fields. 

In the  Fig.(\ref{BS1}) a), the solid black line represents the effective potential without fields, the dashed red line is first energy level, and its probability density is shown in the Fig.(\ref{BS1}) b). We can see that there is only one bound state, and that the electron in this state, is more likely to be found at $\theta=\pi$. When we extend the result, shown by the Fig.(\ref{BS1}) b), for three dimensions, we can visualize a probability ring, whose thickness is $\Delta\theta=0.888\pi$, which is located in the internal region of the toroid. 

In the Figs.(\ref{BS1}) c) and d) the electric field, $E=100$ V/cm is applied, the minimum of the effective potential is shifted to $\theta=\pi/2$. Furthermore the potential well generated under these circumstances is deeper and three bound states are observed. The difference between the first and the second energy level is $0.241$ meV, and the difference between the second and the third energy level is $0.189$ meV. These three energy levels are shown in the Fig.(\ref{BS1}) c) and their probability densities in the Fig.(\ref{BS1}) d). For the first energy level, the ring of probability, whose thickness is $\Delta\theta=0.488\pi$, however, this is located at the top of the toroid. 

For the second energy level, the cloud of probability is in the form of a ring (or probability ring) with two dark stripes, regions most likely to find the electron ($\theta\approx 0.25\pi$ and $\theta\approx 0.8\pi$), interspersed with a light stripe ($\theta\approx 0.5\pi$), the region least likely to find the electron. 

And for the third energy level, the cloud of probability has the form of a ring with three dark stripes, regions most likely to find the electron ($\theta\approx 0$, $\theta\approx 0.5\pi$ and $\theta\approx \pi$), interspersed with two light stripes, the region least likely to find the electron ($\theta\approx 0.3\pi$ and $\theta\approx 0.7\pi$). 

When we apply only the magnetic field, $B=1$ T, the minimum of the effective potential remains at $\theta=\pi$, but unlike the case without field, this potential well is deeper and four more energy levels are confined. The difference between the first and the second energy level is $0.252$ meV, and the difference between the second and the third energy level is $0.239$ meV. The three first energy levels are shown in the Fig.(\ref{BS1}) e), and their the probability densities are show in the Fig.(\ref{BS1}) f). 

For the first energy level, the cloud of probability has the shape of the ring and its thickness is $\Delta\theta=0.488\pi$, and it is located at the internal region of the toroid. 

For the second energy level, the cloud of probability is in the form of a ring with two dark stripes, regions most likely to find the electron ($\theta\approx -0.7\pi$ and $\theta\approx 0.7\pi$), interspersed with a light stripe ($\theta\approx 0$), the region least likely to find the electron.

And for the third energy level, the cloud of probability is in the form of a ring with three dark stripes, regions most likely to find the electron ($\theta\approx -0.52\pi$, $\theta\approx 0.52\pi$ and $\theta\approx \pi$), interspersed with two light stripes, the region least likely to find the electron ($\theta\approx -0.78\pi$ and $\theta\approx 0.78\pi$).

When we combine the magnetic fields, $B=1$ T, and electric, $E=100$ V/cm, we observe the minimum of the effective potential is slightly moved to somewhere between $\theta=\pi/2$ and  $\theta=\pi$. For these conditions five energy levels are confined on the surface of the toroid. The difference between the first and the second energy level is $0.315$ meV, and the difference between the second and the third energy level is $0.280$ meV. The first three energy levels and their probability densities are shown in the Figs.(\ref{BS1}) g) and h). For the first energy level, the cloud of probability has the shape of the ring and its thickness is $\Delta\theta=0.4286\pi$, and it is between the top and the internal region of the toroid, more precisely in $\theta\approx 0.75\pi$.

For the second energy level, the cloud of probability is in the form of a ring with two dark stripes, regions most likely to find the electron ($\theta\approx 0.52\pi$ and $\theta\approx -0.95\pi$), interspersed with a light stripe ($\theta\approx 0.77\pi$), the region least likely to find the electron.

And for the third energy level, the cloud of probability is in the form of a ring with three dark stripes, regions most likely to find the electron ($\theta\approx -0.76\pi$, $\theta\approx 0.37\pi$ and $\theta\approx 0.79\pi$), interspersed with two light stripes, the region least likely to find the electron ($\theta\approx 0.6\pi$ and $\theta\approx \pi$).

\begin{figure*}[h!]
\begin{center}
\includegraphics[scale=0.6]{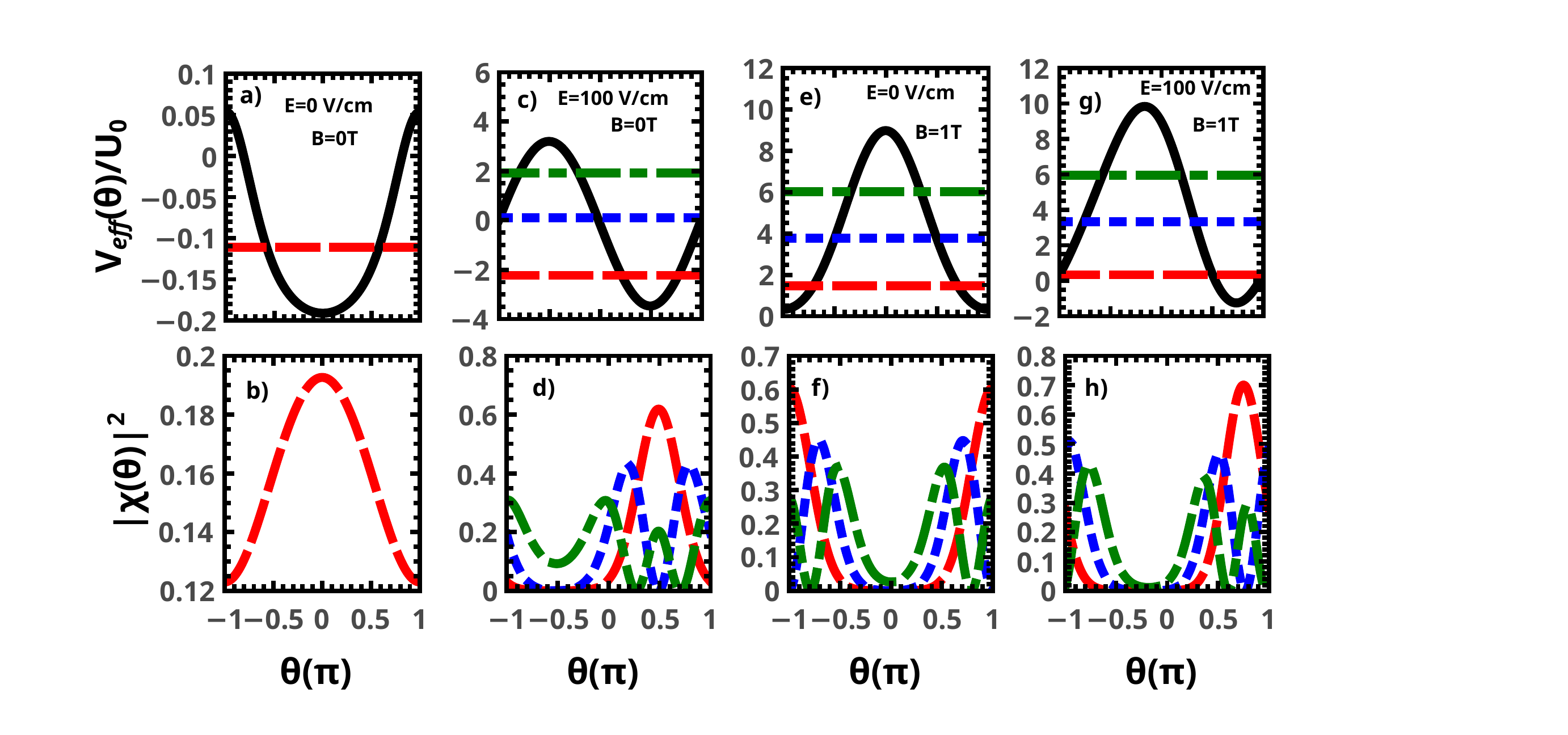}
\caption{Bounds states and the their probability densities for a torus configuration with $r=350$\AA \,\,and $a=900$\AA. The solid black line represents the effective potential, for $m=1$. The dashed red, the dotted blue and the dash-dotted green lines correspond to the first, the second and the third energy levels and their probability densities, respectively.}
\label{BS2}
\end{center}
\end{figure*}

For the same geometry of the Fig.(\ref{BS1}) we investigate the energy levels and the their probability densities for $m=1$. For the case without fields, we can see the Fig.(\ref{BS2}) a), there is only one energy level confined, and an electron in this state is more likely to be found at $\theta\approx 0$, as is shown in the Fig.(\ref{BS2}) b), unlike the case for $m=0$, as we can see in the Fig.(\ref{BS1}) a), where an electron is more likely to be found at $\theta\approx \pi$, see the Fig.(\ref{BS1}) b).

When the electric and magnetic fields are applied to electrons on toroid for the orbital angular momentum, $m=1$, we observe that the manipulation of bound states and how their cloud probability is distributed over the surface of the toroid, via these fields, is similar qualitatively to the case $m=0$, already discussed, compare the Figs.(\ref{BS2}) c), e) and g) with the Figs.(\ref{BS1}) c), e) and g). 

In the Fig.(\ref{BS2}) c), $E=100$ V/cm, the difference between the first and the second energy level is $0.244$ meV, and the difference between the second and the third energy level is $0.189$ meV. 

In the Fig.(\ref{BS2}) e), $B=1$ T, the difference between the first and the second energy level is $0.242$ meV, and the difference between the second and the third energy level is $0.235$ meV.

And, in the Fig.(\ref{BS2}) g), $E=100$ V/cm, $B=1$ T,the difference between the first and the second energy level is $0.313$ meV, and the difference between the second and the third energy level is $0.275$ meV.

According to the results, we observe that the electric field tends to move the cloud of probability to $\theta=\pi/2$, while the magnetic field tends to move it to $\theta=\pi$.

\begin{figure*}[h!]
\begin{center}
\includegraphics[scale=0.6]{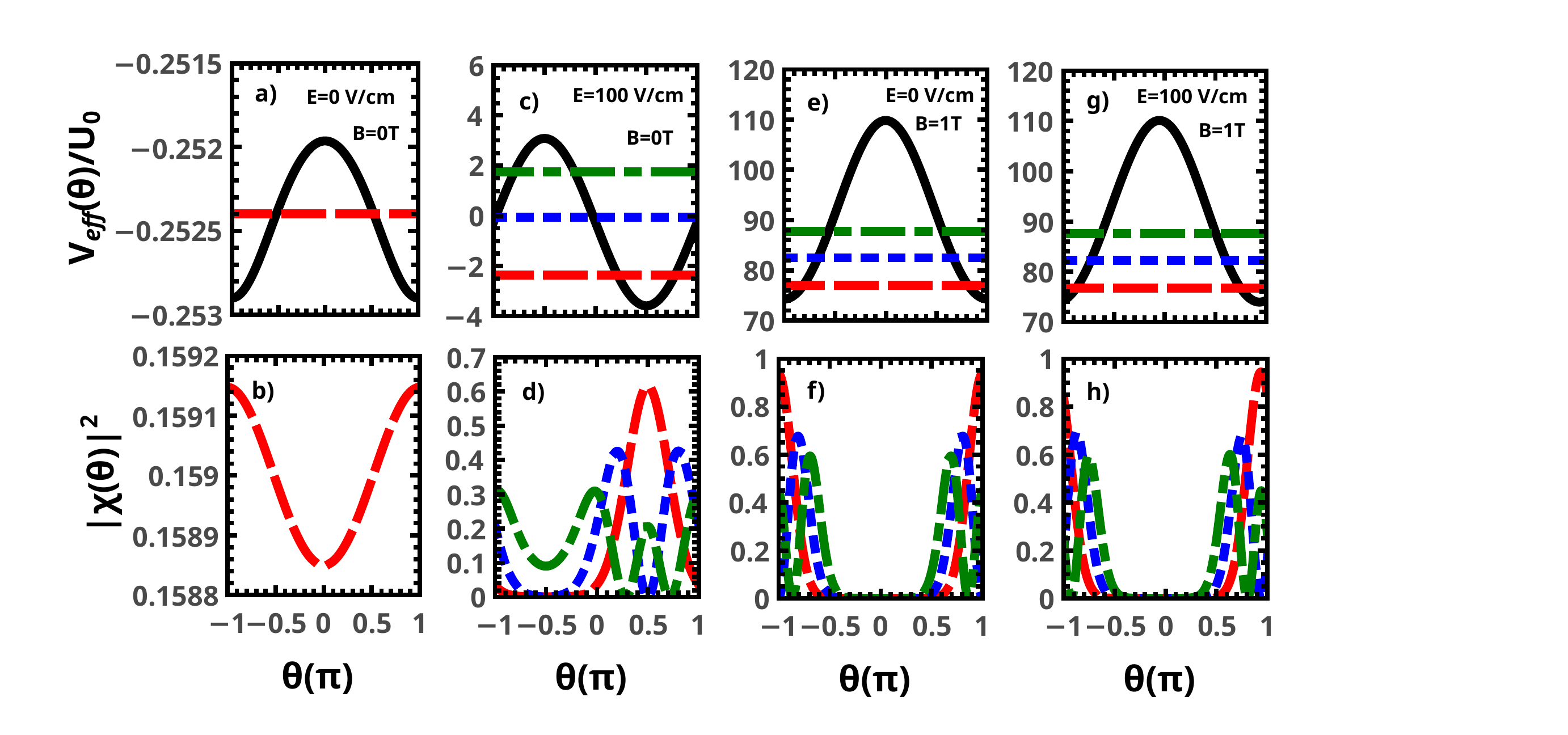}
\caption{Bounds states and the their probability densities for a torus configuration with $r=350$\AA \,\,and $a=3600$\AA. The solid black line represents the effective potential, for $m=0$. The dashed red, the dotted blue and the dash-dotted green lines correspond to the first, the second and the third energy levels and their probability densities, respectively.}
\label{BS3}
\end{center}
\end{figure*}

\begin{figure*}[h!]
\begin{center}
\includegraphics[scale=0.6]{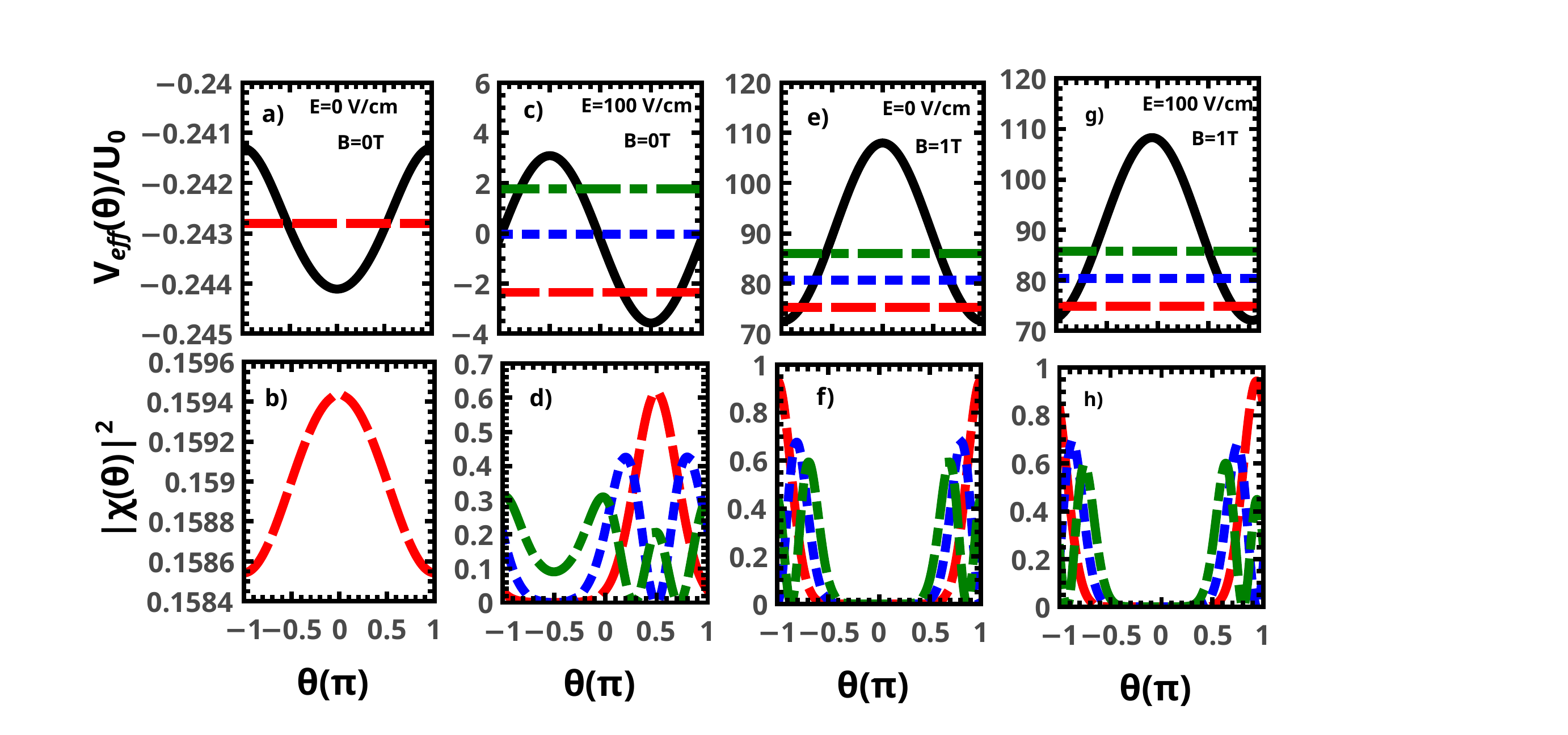}
\caption{Bounds states and the their probability densities for a torus configuration with $r=350$\AA \,\,and $a=3600$\AA. The solid black line represents the effective potential, for $m=1$. The dashed red, the dotted blue and the dash-dotted green lines correspond to the first, the second and the third energy levels and their probability densities, respectively.}
\label{BS4}
\end{center}
\end{figure*}

The energy levels and their probability densities were also studied for a ring-type geometry toroid, $a=3600$ \AA, $r=350$ \AA. In the Fig.(\ref{BS3}) and the Fig.(\ref{BS4}) are shown the energy levels and their probability densities for orbital angular momentum $m=0$ and $m=1$, respectively, under the same field configurations given by the Figs.(\ref{BS1}) and (\ref{BS2}). The cloud of probability of first energy level, for $m=0$ and without field, is a ring located at $\theta=\pi$ wide $\Delta\theta\approx 1.012\pi$, which is less localized than for the geometry of the toroid, $a=900$ \AA, considering that $\Delta\theta\approx 0.888\pi$, see the Figs.(\ref{BS1}) b) and (\ref{BS3}) b) 

The difference between the energy levels, under the action of the electric field, is not affected when the geometry of the toroid changes from $a=900$\AA to $a=3600$ \AA, as shown in the Figs.(\ref{BS3}) c) and (\ref{BS1}) c). 

However, the difference between the energy levels, under the action of the magnetic field is affected with the change of the geometry, for example, the difference between the first and the second energy level is $0.570$ meV, and the difference between the second and the third energy level is $0.555$ meV, i.e., the difference between the levels increases more than twice when the dimension of the toroid increases from $a=900$ to $3600$ \AA, according to Figs.(\ref{BS1}) e) and (\ref{BS3}) e). Therefore, the increase in the difference between the energy levels evidences the fact that the bound states become more confined, this is also shown in the Fig.(\ref{BS3}) f), where the probability densities are narrower (for the first state, for instance, is narrower $\Delta\theta\approx 0.316\pi$.).   

And, in the Figs.(\ref{BS3}) g), $E=100$ V/cm, $B=1$ T, the difference between the first and the second energy level is $0.578$ meV, and the difference between the second and the third energy level is $0.559$ meV, which are 83.49\% and 99.64\%, respectively, greater when comparing the differences between the energy levels of the toroid geometry $a=900$ \AA, see the Fig.(\ref{BS1}) g). The probability densities are shown in the Fig.(\ref{BS3}) h).

The influence of the change in toroid size on the energy levels for $m=0$, described above, are similar for $m=1$, as we see in the Figs.(\ref{BS2}) and (\ref{BS4}).

\section{final remarks}

We study the electronic states of an electron on a graphene toric surface. By writing the Laplacian in the toric coordinate system and considering the da Costa curvature-dependent potential, the effects of the geometry were explored. Moreover, the influence of the external electric and magnetic fields were considered. The axial symmetry with respect to the $z$ axis and the
$\mathcal{PT}$ symmetry allow us to obtain an effective one-dimensional hermitean equivalent Hamiltonian.  The Schrödinger equation is solved numerically, using periodic boundary conditions, since $ V_ {eff}(\theta+2\pi)=V_{eff}(\theta)$.

In absence of external fields, for $m=0$, only one bound state was found located at the internal region the torus, around $\theta=\pi$. For $m\neq0$, the minimum of the potential is shifted to the outer region of the at $\theta=0$, where the electron is localized due to the curvature. The geometric effect of the effective potential increases with the ratio $a/r$, thereby increasing the electron confinement on the toroid surface.

By applying an external electric field, the number of the bound states increases and the minimum of the potential is displayed, forming rings regions where the wave function is concentrated. For instance, assuming $E=100$ V/cm the bound state is located around $\theta=0.5\pi$.



An external magnetic field displaces the minimum of the potential to the inner region, regardless the value of the angular momentum. Furthermore, the number of bound states and their differences increase with the torus radius $a$.

We observe that the electric and magnetic fields compete with each other, for the electric field tends to move the wave function ring to $\theta=\pi/2$ whereas the magnetic field tries to move it to $\theta=\pi$. In addition, by increasing the intensity of the fields, more bound states can be obtained whose spectrum exhibits more separate energy levels. These features show that the states are highly manipulable via external fields, which suggest that electrons on a torus a promising system to a wide range of future technological applications.


Finally, we show that these states are highly manipulable via external fields, which indicates that electrons on the surface of a toroid can be a promising system in a wide range of future technological applications.

\end{document}